\documentclass[letter,twocolumn]{jpsj3}
\usepackage{txfonts}
\usepackage{color}
\usepackage{amsmath}
\usepackage{here}

\title{Cycloidal Magnetic Ordering in Noncentrosymmetric EuIrGe$_3$}

\author{Takeshi Matsumura$^{1}$, Mitsuru Tsukagoshi$^{1}$, Yoshihisa Ueda$^{1}$, Nonoka Higa$^{1}$, Akiko Nakao$^{2}$,  Koji Kaneko$^{3,4}$,   Masashi Kakihana$^{5}$, Masato Hedo$^{5}$,  Takao Nakama$^{5}$, and Yoshichika \={O}nuki$^{5,6}$ }

\inst{$^{1}$Department of Quantum Matter, ADSE, Hiroshima University, Higashi-Hiroshima, Hiroshima 739-8530, Japan \\
$^{2}$Comprehensive Research Organization for Science and Society, Tokai, Ibaraki 319-1106, Japan \\
$^{3}$Materials Sciences Research Center, Japan Atomic Energy Agency, Tokai, Ibaraki, 319-1195, Japan \\
$^{4}$J-PARC Center, Japan Atomic Energy Agency, Tokai, Ibaraki, 319-1195, Japan \\
$^{5}$Faculty of Science, University of the Ryukyus, Nishihara, Okinawa 903-0213, Japan\\
$^{6}$RIKEN Center for Emergent Matter Science, Wako, Saitama 351-0198, Japan\\
}

\abst{
Successive magnetic phase transitions at $T_{\text{N}}$=12.2 K, $T_{\text{N}}^{\;\prime}$=7.0 K, and $T_{\text{N}}^{\;*}$=5.0 K in EuIrGe$_3$, an intermetallic compound with a body centered tetragonal lattice belonging to a polar space group $I4mm$, has been investigated by neutron diffraction and resonant X-ray diffraction. 
It is shown that EuIrGe$_3$ exhibits an incommensurate longitudinal sinusoidal order with $\mib{q}\sim (0, 0, 0.792)$ and $\mib{m}_{\mib{q}} \parallel c\text{-axis}$ in the high temperature phase ($T_{\text{N}}^{\;\prime}< T < T_{\text{N}}$), which changes to a cycloidal order with $\mib{q}=(\delta', 0, 0.8)$ ($\delta'\sim 0.017$) and $\mib{m}_{\mib{q}} \parallel ac\text{-plane}$ in the intermediate phase ($T_{\text{N}}^{\;*} < T < T_{\text{N}}^{\;\prime}$). 
In the low temperature phase ($T < T_{\text{N}}^{\;*}$), the cycloidal plane rotates by $45^{\circ}$ to have $\mib{q}=(\delta, \delta, 0.8)$ ($\delta\sim 0.012$). 
It is also pointed out that the X-ray scattering amplitude from odd-parity magnetic quadrupole due to the polar environment interfere with that from normal even-parity magnetic dipole in the magnetic ordered phase. 
(\today)
}

\begin{document}
\maketitle

In noncentrosymmetric magnetic materials, various kinds of nontrivial magnetic structures are realized as a result of competing interactions of symmetric magnetic exchange interaction, Dzyaloshinskii-Moriya (DM) antisymmetric exchange interaction, and the Zeeman energy in external fields.~\cite{Dzyaloshinsky58,Moriya60}  
In many cases, they are non-collinear or non-coplanar structures associated with incommensurate spiral orderings. 
Typical examples are the formation of skyrmion lattices in cubic chiral magnets such as MnSi with the space group $P2_13$.~\cite{Muhlbauer09,Neubauer09} 
A similar skyrmion lattice phase is also observed in a rare-earth chiral magnet EuPtSi, which belongs to the same space group.~\cite{Kakihana18,Kakihana19,Kaneko19,Tabata19,Homma19,Sakakibara19,Takeuchi19} 
Recently, skyrmion lattices have also been discovered even in centrosymmetric Gd compounds.~\cite{Kurumaji19,Hirschberger20,Hirschberger19,Khanh22} 
The $S$=7/2 ($L$=0) total spin state in Eu$^{2+}$ and Gd$^{3+}$ provides an ideal platform to investigate the formations of characteristic spin structures which are little affected by the crystal-field anisotropy.  

\begin{figure}
\begin{center}
\includegraphics[width=8.5cm]{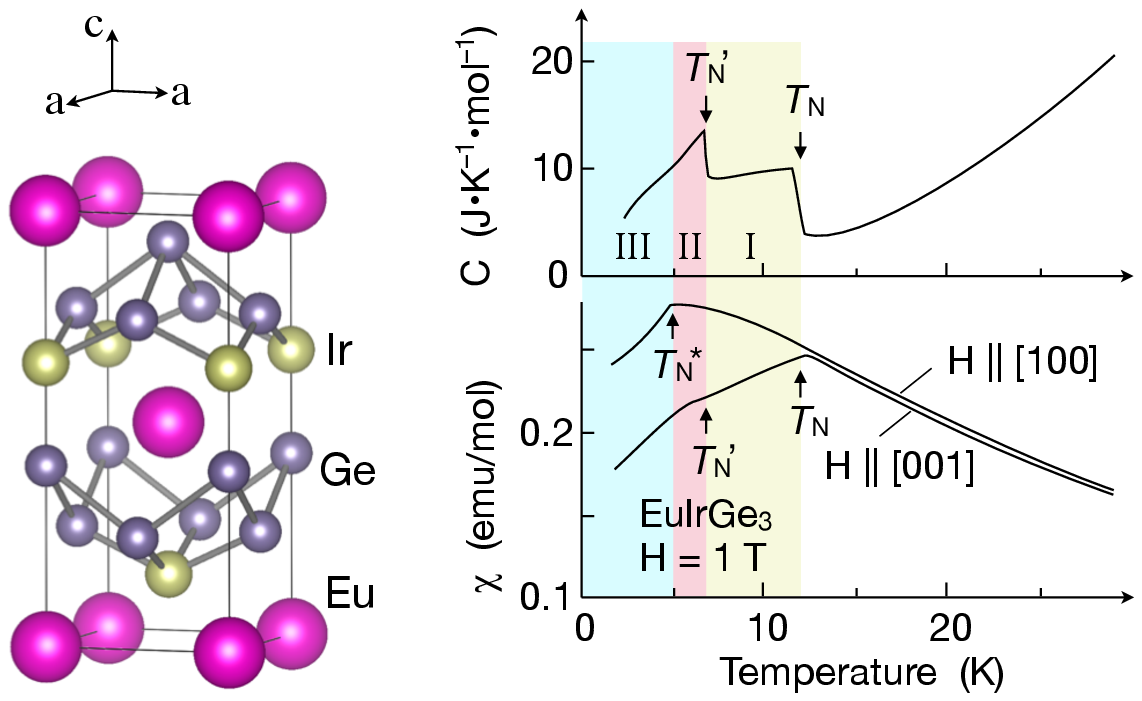}
\caption{(Color online) left: Crystal structure of EuIrGe$_3$ drawn with the program VESTA.~\cite{Momma11} $a=4.436$ \AA\ and $c=10.053$ \AA\ at room temperature.~\cite{Kakihana17} 
right: Temperature dependences of specific heat and magnetic susceptibilities reproduced from the literature.~\cite{Kakihana17} 
 }
\label{fig:Structure}
\end{center}
\end{figure}

EuTGe$_3$ (T=transition metal), with the BaNiSn$_3$-type body-centered tetragonal structure (space group $I4mm$), exhibits a wide variety of magnetic properties.~\cite{Kumar12,Kaczorowski12,Kakihana17,Rai21} 
In EuNiGe$_3$, the helimagnetic order and the appearance of anomalous Hall effect in the intermediate phase in magnetic field suggest a formation of a skyrmion lattice state.~\cite{Goetsch13,Maurya14,Ryan16,Fabreges16,Iha20} 
 The magnetic properties of EuRhGe$_3$ and EuIrGe$_3$ have also been well studied.~\cite{Bednarchuk15,Maurya16,Utsumi18} 
In spite of weak magnetic anisotropies of Eu$^{2+}$, these compounds exhibit multiple magnetic phases in the temperature ($T$) vs magnetic field ($H$) space, suggesting a subtle energy balance among various magnetic structures, which is of great interest. 
EuIrGe$_3$, the subject compound of this paper, exhibits successive phase transitions at $T_{\text{N}}$=12.2 K,  $T_{\text{N}}^{\;\prime}$=7.0 K, and $T_{\text{N}}^{\;*}$=5.0 K.~\cite{Maurya16,Kakihana17}
The crystal structure and the physical properties of EuIrGe$_3$ are summarized in Fig.~\ref{fig:Structure}, where the three magnetic phases at zero field are named as phase I, II, and III, respectively.  
However, the detailed magnetic structures of these phases have not been clarified yet. 
From the structural viewpoint, it is of special interest that the crystal lacks an inversion center but possesses vertical mirror planes including the $c$-axis (point group $C_{4v}$), allowing the existence of a DM interaction. This is not a chiral, but a polar system. 
In this paper, we report the results of neutron and resonant x-ray diffraction and present the magnetic structures of EuIrGe$_3$ at zero field. 

Single crystals of EuIrGe$_3$, using natural Eu, were grown by the In-flux method as described in Ref.~\citen{Kakihana17}. 
Neutron diffraction was performed at BL18 (SENJU) time-of-flight single crystal neutron diffractometer at the Materials and Life Science Experimental Facility (MLF) of Japan Proton Accelerator Research Complex (J-PARC).~\cite{Ohhara16} The wavelength range of the incident neutrons was $0.4 \sim 4.4$ \AA. 
A plate-shaped single crystal with a $c$-plane surface, $1.5\times 2.0$ mm$^2$ in area and 0.4 mm in thickness, was mounted in a closed cycle $^4$He refrigerator so that the $c$-axis was perpendicular to the incident beam in the horizontal $(H0L)$ scattering plane with the $b$-axis vertical. The crystal was then rotated by $\sim 15^{\circ}$ about the $b$-axis to increase the irradiation area and collect Bragg peaks as many as possible in the reflection condition. 
Neutron diffraction patterns were collected between 4 K and 15 K using a closed cycle $^4$He refrigerator. 

Resonant X-ray diffraction (RXD) was performed at BL-3A of the Photon Factory, KEK, Japan. 
The single crystalline sample was spark cut into a plate-shape and the $c$-plane surface was polished to a shining surface. 
The measurement was performed in the $(H0L)$ scattering plane. The X-ray energy around the Eu $L_2$-edge was used. 
See supplemental material (SM) for the scattering configuration of the RXD experiment.~\cite{SM} 

\begin{figure}
\begin{center}
\includegraphics[width=8.5cm]{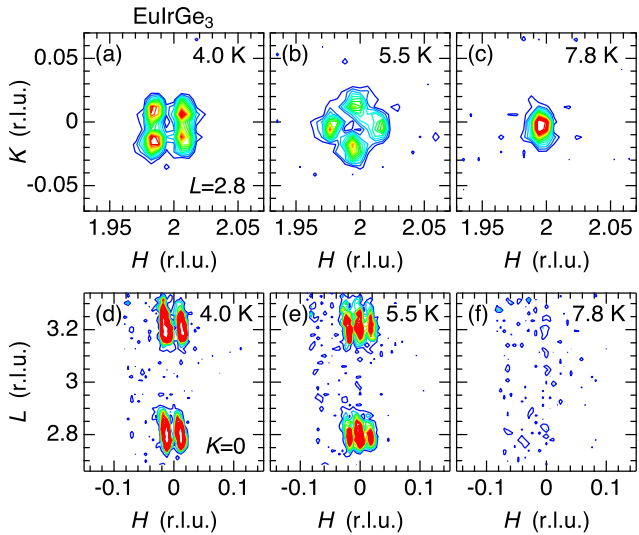}
\caption{(Color online) Contour maps of neutron-diffraction intensity around (a--c) (2, 0, 2.8) and (d--f) (0, 0, 3) at 4.0 K, 5.5 K, and 7.8 K in phases III, II, and I, respectively. 
 }
\label{fig:Map1}
\end{center}
\end{figure}

In the neutron diffraction experiment at SENJU, we successfully detected magnetic signals appearing below $T_{\text{N}}$ at positions around the propagation vector corresponding to $\mib{q}=(0, 0, 0.8)$. 
Figure~\ref{fig:Map1} shows the contour maps of neutron-diffraction intensity around (2, 0, 2.8) and (0, 0, 3), demonstrating the splitting of the Bragg peak below $T_{\text{N}}^{\;\prime}$ and $T_{\text{N}}^{\;*}$. 
The $(H, K, 0)$ cut of Fig.~\ref{fig:Map1}(a) and the $(H, 0, L)$ cut of Fig.~\ref{fig:Map1}(d) clearly demonstrate that the magnetic propagation vector at 4.0 K in phase III is described by $\mib{q}_3=(\delta, \delta, 0.8)$ with $\delta \sim 0.012$. 
The four peaks are considered to be due to the magnetic domains with equivalent $\mib{q}$-vectors. 
At 5.5 K in phase II, as demonstrated in Figs.~\ref{fig:Map1}(b) and \ref{fig:Map1}(e), the positions of the four peaks rotate by $45^{\circ}$, indicating that the propagation vector in phase II is  described by $\mib{q}_2=(\pm\delta', 0, 0.8)$ with $\delta' \sim 0.017$. 
At 7.8 K in phase I, the four peaks merge to a single peak as shown in Fig.~\ref{fig:Map1}(c), indicating that the magnetic structure of phase I is described by $\mib{q}_1=(0, 0, 0.8)$. Since the peak disappears on the $(0, 0, L)$ line as shown in Fig.~\ref{fig:Map1}(f), we can conclude that the moments are aligned along the $c$-axis in phase I. 

\begin{figure}
\begin{center}
\includegraphics[width=8.5cm]{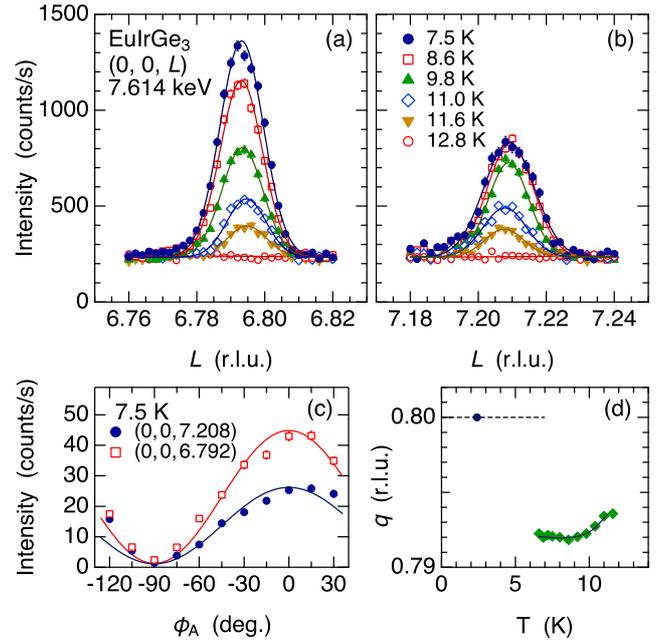}
\caption{(Color online) (a) and (b): Temperature dependence of the reciprocal space scan along $(0, 0, L)$ around $L=6+q$ and $L=8-q$ ($q\sim 0.8$), respectively, at the $E1$ resonance energy of 7.614 keV without polarization analysis. 
Solid lines are the fits with Gaussian functions. 
(c) Polarization analysis of the resonant Bragg peak at $(0, 0, 7.208)$ and $(0, 0, 6.792)$ ($q=0.792$). 
The intensity at $\phi_{\text{A}}=-90^{\circ}$ and $0^{\circ}$ corresponds to the $\pi$-$\pi'$ and $\pi$-$\sigma'$ intensity, respectively. 
(d) Temperature dependence of the $q$ value along $L$ in phase I. 
}
\label{fig:Lscanspolq}
\end{center}
\end{figure}

The magnetic Bragg peaks have been investigated in more detail by RXD with higher spatial resolution. 
Figures~\ref{fig:Lscanspolq}(a) and \ref{fig:Lscanspolq}(b) show the results of $(0, 0, L)$-scans around $L=6.8$ and 7.2 with increasing $T$ in phase I. 
These data clearly demonstrate that the peak position slightly deviates from the commensurate value at $q=0.8$. The magnetic propagation vector is more precisely described by $\mib{q}_1=(0, 0, 0.792)$. Furthermore, the fitting result shows that $q$ is weakly dependent on $T$ as shown in Fig.~\ref{fig:Lscanspolq}(d). 
The $L$-scan at 2.4 K in phase III shows that the peak is exactly at $q=0.8$, which is presented in the SM with the result of the energy spectrum of the resonance.~\cite{SM}  

The result that the ordered moments in phase I are parallel to the $c$-axis is also confirmed by the polarization analysis in RXD, which is demonstrated in Fig.~\ref{fig:Lscanspolq}(c). 
If the Fourier component of the magnetic structure had finite in-plane component along the $b$-axis, which is perpendicular to the scattering plane in the present configuration, there should arise finite $\pi$-$\pi'$ intensity at $\phi_{\text{A}}=-90^{\circ}$. The vanishing intensity of $\pi$-$\pi'$ indicates that the moments are parallel to the $c$ axis. 
The solid lines are the calculations for the magnetic scattering at the $E1$ resonance, i.e., $F \propto (\mib{\varepsilon}' \times \mib{\varepsilon} ) \cdot \mib{m}$,~\cite{Hannon88,Blume94,Lovesey05} by assuming the longitudinal sinusoidal magnetic structure with the $c$-axis component only.

\begin{figure}
\begin{center}
\includegraphics[width=8.5cm]{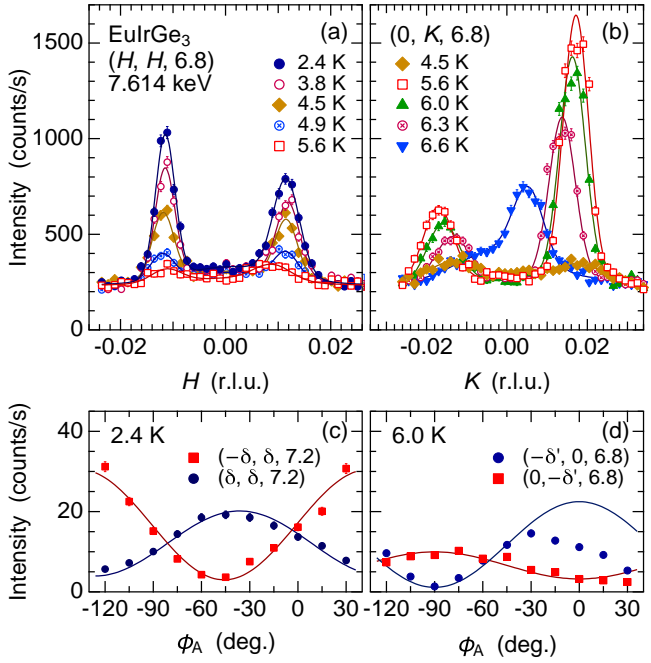}
\caption{(Color online) (a) and (b): Temperature dependence of the reciprocal space scan along $(H, H, 6.8)$ and $(0, K, 6.8)$, respectively, at the $E1$ resonance energy of 7.614 keV without polarization analysis. Solid lines are the fits with Gaussian functions. 
(c) and (d): Polarization analysis of the resonant Bragg peaks at 2.4 K in phase III ($\delta=0.012$) and at 6.0 K in phase II ($\delta'=0.017$), respectively. 
The solid lines are the calculated curves for the magnetic scattering at the $E1$ resonance energy by assuming the cycloidal magnetic structure as described in the text and shown in Figs.~\ref{fig:Magst}(b) and (c). 
}
\label{fig:HH0Kpol}
\end{center}
\end{figure}

The reciprocal space scans of $(H, H, 6.8)$ in phase III and  $(0, K, 6.8)$ in phase II at several temperatures are shown in Fig.~\ref{fig:HH0Kpol}(a) and \ref{fig:HH0Kpol}(b), respectively. 
As shown in Fig.~\ref{fig:Map1}, the $(H, H, 6.8)$ scan in phase III exhibits peaks at $H=\pm 0.012$. 
The intensity decreases with increasing $T$ and disappears on entering phase II, leaving weak peaks at 5.6 K. 
In the $(0, K, 6.8)$ scan, on the other hand, the peaks at $K=\pm 0.017$ are very weak in phase III at 4.5 K and develop above 5 K on entering phase II. 
The intensity soon begins to decrease with increasing $T$, and the peak position moves to the central position at $(0, 0, 6.8)$. 
The results of the scans along $(H, -H, 6.8)$ and $(H, 0, 6.8)$ are provided in the SM.~\cite{SM} 

The polarization analyses performed on the representative peaks at $(\delta, \delta, 7.2)$ and $(-\delta, \delta, 7.2)$ ($\delta=0.012$) at 2.4 K in phase III and at $(-\delta', 0, 6.8)$ and $(0, -\delta', 6.8)$ ($\delta'=0.017$) at 6.0 K in phase II are shown in Figs.~\ref{fig:HH0Kpol}(c) and \ref{fig:HH0Kpol}(d), respectively. 
It is clearly demonstrated that the intensity for $(\delta, \delta, 7.2)$ takes the maximum at an intermediate angle around $\phi_{\text{A}}\sim -45^{\circ}$. 
The phase of the $\phi_{\text{A}}$ dependence is reversed for $(-\delta, \delta, 7.2)$, where the intensity takes the minimum at around $\phi_{\text{A}}\sim -45^{\circ}$. 
At 6.0 K in phase II, where the peak position is rotated by $45^{\circ}$, the intensity for $(-\delta', 0, 6.8)$ takes the minimum and maximum at $\phi_{\text{A}}=-90^{\circ}$ ($\pi'$) and $0^{\circ}$ ($\sigma'$), respectively. The decrease in intensity at $-15^{\circ} \le \phi_{\text{A}} \le 30^{\circ}$ is considered to be due to some misalignment of the detector that unexpectedly happened when rotating the analyzer system. 
The phase relation is reversed for $(0, -\delta', 6.8)$, where the intensity is maximum and minimum at $\phi_{\text{A}}=-90^{\circ}$ and $0^{\circ}$, respectively.

\begin{figure}
\begin{center}
\includegraphics[width=8cm]{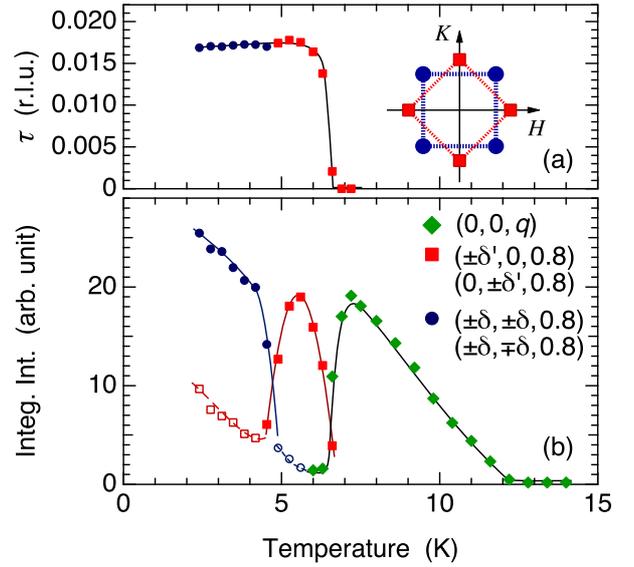}
\caption{(Color online) (a) Temperature dependence of $\tau=\delta'$ in phase II and $\tau=\sqrt{2} \delta$ in phase III.  
(b) Temperature dependence of the total integrated intensity of the resonant Bragg peaks of $(\pm\delta, \pm\delta, q)$ and $(\pm\delta, \mp\delta, q)$ in phase III (closed circles), 
$(\pm\delta', 0, q)$ and $(0, \pm\delta', q)$ in phase II (closed squares), and $(0, 0, q)$ in phase I (closed diamonds), which are obtained from $(H, H, 6.8)$ and $(H, -H, 6.8)$ scans in phase III, $(H, 0, 6.8)$ and $(0, K, 6.8)$ scans in phase II, and $(0, 0, L)$ scans in phase I, respectively. 
The open squares and circles represent the integrated intensity of the residual peaks for the $(H, 0, 6.8)$ and $(0, K, 6.8)$ scans in phase III, and $(H, H, 6.8)$ and $(H, -H, 6.8)$ scans in phase II, respectively, which detect the intensities when crossing the ridges.  
The inset in (a) represents the four peaks and the ridge in phase II (squares) and phase III (circles).  
}
\label{fig:TdepQInt}
\end{center}
\end{figure}

The $T$-dependence of $\delta'$ in phase II and $\delta\sqrt{2}$ in phase III, the distance ($\tau$) from the central position at $(0, 0, 0.8)$, is shown in Fig.~\ref{fig:TdepQInt}(a). This figure shows that $\tau$ does not change at the II--III phase boundary; i.e., the peak position only rotates by $45^{\circ}$. 
Figure~\ref{fig:TdepQInt}(b) shows the $T$-dependence of the total integrated intensity. 
Note that the $(H, 0, 6.8)$ and $(0, K, 6.8)$ scans in phase III, where the peaks of phase II do not exist, pass across the ridge connecting the four peaks in phase III. This gives rise to residual peaks. The intensity is plotted by the open marks. 
The $T$-dependence of the total intensity in Fig.~\ref{fig:TdepQInt}(b) shows that the magnitude of the ordered moment increases with decreasing $T$ below $T_{\text{N}}$. The almost linear $T$-dependence suggest a mean-field-like development: $m \propto \sqrt{T_{\text{N}} - T}$.   
Below $T_{\text{N}}^{\;\prime}$ and $T_{\text{N}}^{\;*}$, although the intensity ratio among the four peaks is not equal due to unequal domain population, the total intensity is maintained and keeps increasing with decreasing $T$.

The magnetic moment of Eu on the $j$th lattice point at $\mib{r}_j$ is generally expressed as
\[
\mib{\mu}_j = \mib{m}_{\mib{q}} e^{i\mib{q}\cdot\mib{r}_j} + \mib{m}_{\mib{q}}^* e^{-i\mib{q}\cdot\mib{r}_j}\;,
\]
where $\mib{m}_{\mib{q}}$ represents the Fourier component of the magnetic structure. 
It is usually reasonable to consider that $\mib{m}_{\mib{q}}$ belongs to one of the irreducible representations of the magnetic structure described by the propagation vector $\mib{q}$ in the space group $I4mm$.~\cite{SM} 
For $\mib{q}=(0, 0, \zeta)$ $(0 < \zeta < 1)$, $\mib{m}_{\mib{q}}$ can be $(0,0,1)$ or a linear combination of $(1,1,0)$ and $(1,\bar{1},0)$. 
The experiment shows $\mib{m}_{\mib{q}}$ = $(0, 0, 1)$ is realized. 

For $\mib{q}=(\delta, \delta, q)$, $\mib{m}_{\mib{q}}$ can be $(1,\bar{1},0)$ or a linear combination of $(0,0,1)$ and $(1,1,0)$. 
The calculated curves of the polarization analysis for $(\delta, \delta, 7.2)$ and $(-\delta, \delta, 7.2)$ in Fig.~\ref{fig:HH0Kpol}(c) are obtained by assuming $\mib{m}_{\mib{q}}$ = $(1/\sqrt{2}, 1/\sqrt{2}, \pm i)$ and $(-1/\sqrt{2}, 1/\sqrt{2}, \pm i)$, respectively, with equal $c$-axis and $ab$-plane amplitudes. 
This indicates the cycloidal propagation of the ordered moment along the $[1, 1, 0]$ and $[1, \bar{1}, 0]$ direction, respectively, and also along the $c$-axis. 
In the same manner, the calculated curves for $(-\delta', 0, 6.8)$ and $(0, -\delta', 6.8)$ in Fig.~\ref{fig:HH0Kpol}(d) are obtained by assuming $\mib{m}_{\mib{q}}$ = $(1, 0, \pm i)$ and $(0, 1, \pm i)$, respectively. This again shows the cycloidal propagation of the magnetic moment. 
Although the sign of $\pm i$ of $m_{\mib{q}, z}$ represent the sense of rotation, we cannot distinguish the two possibilities in this linear polarization analysis. 
This should be checked in future by using a circularly polarized X-ray, 
since the DM vector perpendicular to the mirror plane should favor a unique sense of rotation. 
Also, we presumed the single-$\mib{q}$ structure since the intensities of the four peaks were much different. 
The appearance of the four peaks is ascribed to the domain formation and not to the multi-$\mib{q}$ structure.
The vanishing intensity in the polarization analysis at specific analyzer angles and the good agreement with the calculation well support the single-$\mib{q}$ scenario.

\begin{figure}
\begin{center}
\includegraphics[width=8.5cm]{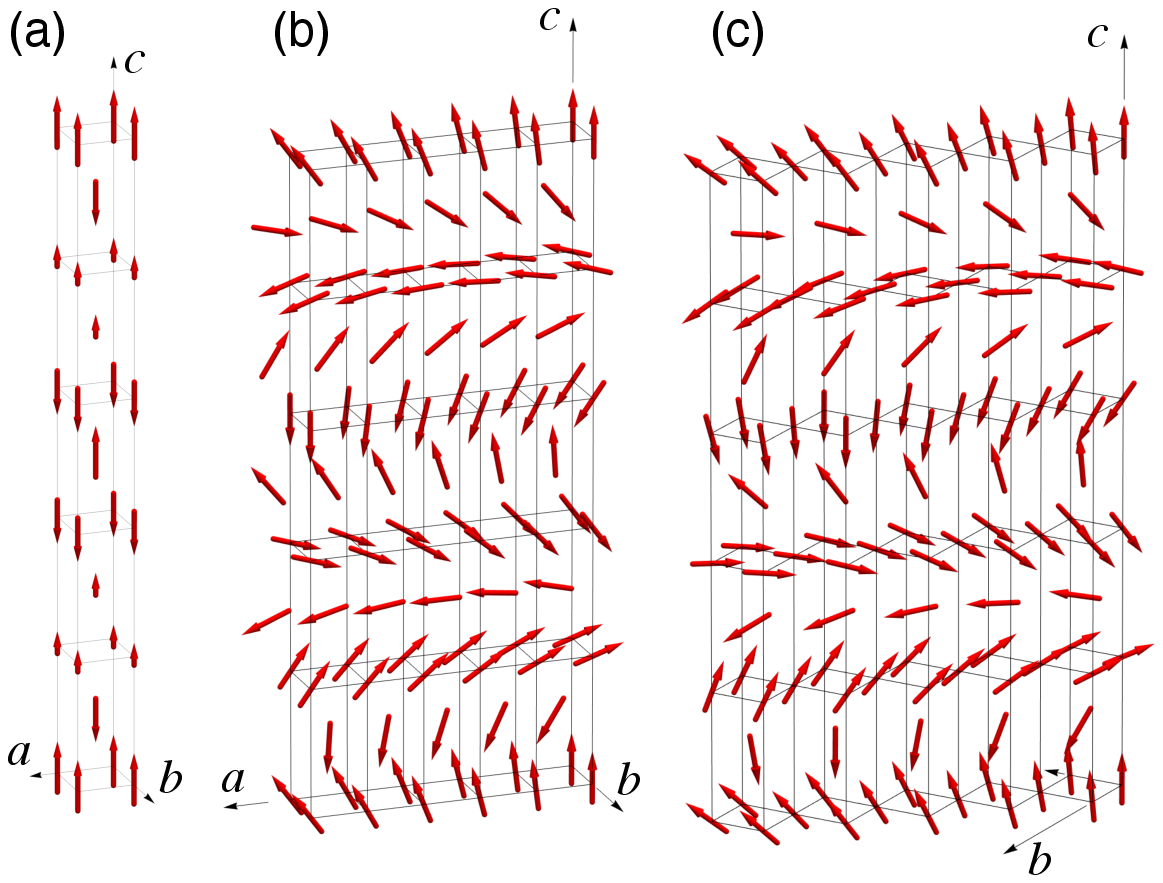}
\caption{(Color online) Models of the incommensurate magnetic structures of EuIrGe$_3$: (a) longitudinal sinusoidal structure with $\mib{q}=(0, 0, \sim 0.792)$ in phase I, (b) cycloidal structure with $\mib{q}=(\delta', 0, 0.8)$ ($\delta'=0.017$) in phase II, and (c) cycloidal structure with $\mib{q}=(\delta, \delta, 0.8)$ ($\delta=0.012$) in phase III. 
}
\label{fig:Magst}
\end{center}
\end{figure}

The magnetic structure of EuIrGe$_3$ in phase I, II, and III is illustrated in Figs.~\ref{fig:Magst}(a), \ref{fig:Magst}(b), and \ref{fig:Magst}(c), respectively. 
By taking a rotation or a mirror reflection of these structures, i.e., by performing symmetry operations allowed in the space group $I4mm$, we obtain the magnetic structures of other domains. 
Just below $T_{\text{N}}=12.2$ K, the longitudinal sinusoidal structure of Fig.~\ref{fig:Magst}(a) is realized. 
The alignment of the ordered moments along the $c$-axis could be associated with the weak magnetic anisotropy to prefer the $c$-axis orientation. 
However, since there inevitably remain small moment sites in this structure, it is not preferable to maintain this structure down to the lowest temperature due to the requirement of entropy reduction and the energy gain in the magnetic exchange interaction. 
It is necessary to increase the size of the magnetic moments. 
In the $L=0$ system with weak anisotropy, this can be realized by giving rise to an in-plane magnetic component. 
Since $q\sim 0.8$ is almost fixed by the Ruderman-Kittel-Kasuya-Yoshida (RKKY) interaction to maximize $J(\mib{q})$, which is associated with the conduction electron state, a possible solution is to rotate the magnetic moment to form a helical or cycloidal structure. 

No change in the crystal structure has been detected experimentally, although more careful measurement would be necessary. 
Within the framework of the $I4mm$ space group, the slight tilt of the $\mib{q}$-vector from $(0, 0, q)$ to $(\delta', 0, q)$ or $(\delta, \delta, q)$ in phase II or III, respectively, could be a consequence of maximizing the energy gain of the magnetic exchange interaction without distorting the tetragonal lattice. 
It is possible to form the cycloidal structure in the $I4mm$ symmetry by tilting $\mib{q}$ since the $c$-axis and the $ab$-plane components of $\mib{m}_{\mib{q}}$ have the same irreducible representation, whereas the cycloid with $\mib{q}=(0, 0, q)$ breaks the symmetry of $I4mm$.~\cite{SM} 

The $45^{\circ}$ rotation of the $\mib{q}$-vector about the $c^*$ axis is also observed in the multiple magnetic transitions in EuAl$_4$.~\cite{Kaneko21,Shimomura19,Shang21} 
It is intriguing that the sequence is opposite to the present case in EuIrGe$_3$; in EuAl$_4$, the $\mib{q}_2=(\delta_2, \delta_2, 0)$ ($\delta_2=0.085$) appears at high $T$ and  $\mib{q}_1=(\delta_1, 0, 0)$ ($\delta_1=0.194$) is realized at low $T$.  
The cycloidal structure at zero field in EuIrGe$_3$ with the $C_{4v}$ crystal class also suggests a possibility for the N\'{e}el type skyrmion formation as observed in a polar 3d system GaV$_4$S$_8$.~\cite{Kezsmarki15} 

Finally, we point out that the resonant intensities at $(0, 0, G+q)$ and $(0, 0, G-q)$ in phase I, where $G=2n$ represents the fundamental Bragg point, are different as shown in Fig.~\ref{fig:Lscanspolq} and in the SM in more detail.~\cite{SM} 
This cannot be explained if we only consider the $E1$ resonance due to the magnetic dipole, where the $\pi$-$\sigma'$ intensity should be proportional to $\sin^2 \theta$. 
The difference in intensity is intrinsic because the intensities of the $(0, 0, G+q)$ and $(0, 0, G-q)$ reflections respectively follow the $\sin^2 \theta$ dependence. 
From the viewpoint of symmetry, in the $I4mm$ space group without a horizontal mirror plane perpendicular to the $c$ axis, the propagation vectors of $(0, 0, q)$ and $(0, 0, -q)$ are different. 
Therefore, the intensities of $(0, 0, G+q)$ and $(0, 0, G-q)$ can be different as experimentally observed. 

This difference can be explained by considering the noncentrosymmetric polar environment at the Eu site as it is apparent in Fig.~\ref{fig:Structure}. 
The finite electric field gradient at the Eu site induces the mixing between $4f$ and $5d$ states, allowing the $E1$-$E2$ cross-term resonance.~\cite{Lovesey05}
The magnetic moment distribution in the hybrid orbital state is anisotropic and should be different at Eu sites with the upward and downward magnetic moments as schematically illustrated in the SM.~\cite{SM} 
By symmetry analysis, this hybrid state is described by a superposition of magnetic dipole, magnetic quadrupole, and magnetic monopole.~\cite{Hayami18a,Hayami18b,Yatsushiro21}  
The contribution from magnetic monopole can be excluded here since the rank-0 quantity does not contribute to the $E1$-$E2$ resonance.~\cite{Lovesey05} 
The scattering amplitudes of normal $E1$-$E1$ resonance due to the magnetic dipole and the $E1$-$E2$ cross-term resonance due to the magnetic quadrupole interfere. 
Then, the structure factors at $(0, 0, G+q)$ and $(0, 0, G-q)$ are expressed as $F_{\text{md}}\alpha_{11}+F_{\text{mq}}\alpha_{12}$ and $F_{\text{md}}^{\;*}\alpha_{11}+F_{\text{mq}}^{\;*}\alpha_{12}$, respectively, where $F_{\text{md}}$ and $F_{\text{mq}}$ are the magnetic dipole and quadrupole structure factors, $\alpha_{11}$ and $\alpha_{12}$ are the $E1$-$E1$ and $E1$-$E2$ spectral functions, respectively. 
This relation leads to the different intensities at $(0, 0, G\pm q)$.~\cite{SM}  
More detailed study to extract the signal from the magnetic quadrupole is necessary. 

In summary, we have investigated the magnetic structures of the three successive magnetic phases of noncentrosymmetric EuIrGe$_3$ by neutron and resonant X-ray diffraction. 
With decreasing temperature, a longitudinal sinusoidal structure with $\mib{q}=(0, 0, 0.792)$ takes place below $T_{\text{N}}$=12.2 K, 
which changes to a cycloidal structure with $\mib{q}=(\delta', 0, 0.8)$ ($\delta'$=0.017) below $T_{\text{N}}^{\;\prime}$=7.0 K. 
The cycloidal plane rotates by $45^{\circ}$ below $T_{\text{N}}^{\;*}$=5.0 K to have $\mib{q}=(\delta, \delta, 0.8)$ ($\delta$=0.012). 
We also detected the $E1$-$E1$ and $E1$-$E2$ interference term due to the parity-odd magnetic quadrupole induced in the $4f$-$5d$ hybrid orbital state, reflecting the lack of inversion symmetry at the Eu site.

\paragraph{Acknowledgements} 
The authors acknowledge valuable discussion with A. Tanaka. 
This work was supported by JSPS KAKENHI Grant number JP20H01854. 
The neutron experiment at the Materials and Life Science Experimental Facility of the J-PARC was performed under a user program (Proposal No. 2020A0118).
The synchrotron experiments were performed under the approval of the Photon Factory Program Advisory Committee (No. 2020G034).

\clearpage
\renewcommand{\topfraction}{1.0}
\renewcommand{\bottomfraction}{1.0}
\renewcommand{\dbltopfraction}{1.0}
\renewcommand{\textfraction}{0.01}
\renewcommand{\floatpagefraction}{1.0}
\renewcommand{\dblfloatpagefraction}{1.0}
\setcounter{topnumber}{5}
\setcounter{bottomnumber}{5}
\setcounter{totalnumber}{5}
\setcounter{page}{1}

\renewcommand{\theequation}{S\arabic{equation}}
\renewcommand{\thefigure}{S\arabic{figure}}
\renewcommand{\thetable}{S-\Roman{table}}
\setcounter{section}{19}
\setcounter{figure}{0}


\begin{fullfigure}[t]
\begin{center}
\textbf{\large{Supplemental Material}}
\end{center}
\vspace{5mm}

\begin{center}
\textbf{\large{Cycloidal Magnetic Ordering in Noncentrosymmetric EuIrGe$_3$}} \\
\vspace{2mm}
Takeshi Matsumura, Mitsuru Tsukagoshi, Yoshihisa Ueda, Nonoka Higa, Akiko Nakao,  Koji Kaneko,   Masashi Kakihana, Masato Hedo,  Takao Nakama, and Yoshichika \={O}nuki
\end{center}
\vspace{20mm}

\begin{minipage}[b]{0.47\textwidth}
\begin{center}
\includegraphics[width=8.5cm]{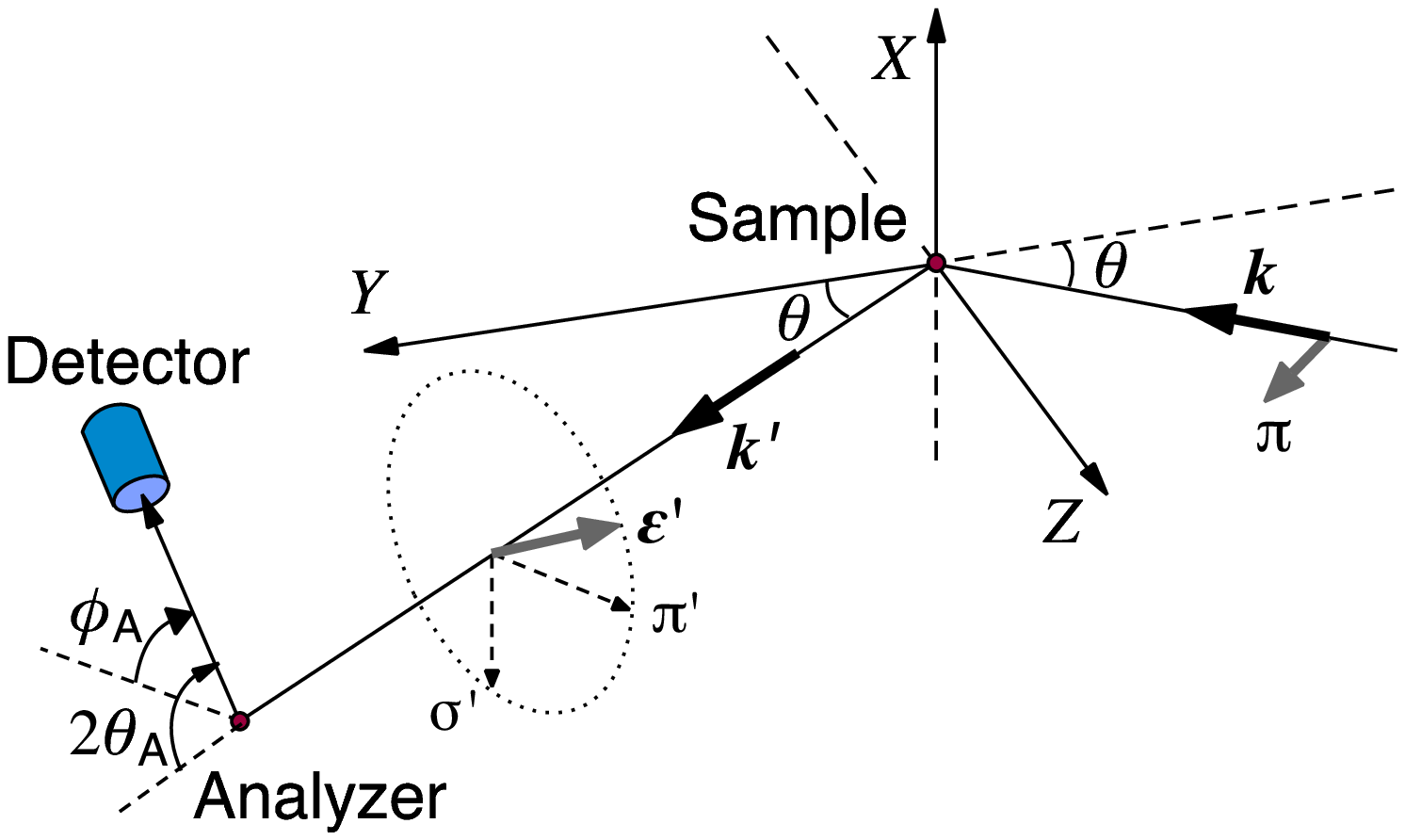}
\end{center}
\caption{(Color online) Scattering configuration of the resonant X-ray diffraction experiment. We take the $Z$-axis along the scattering vector $\mib{Q}=\mib{k}'-\mib{k}$, $X$-axis along $\mib{k}\times\mib{k}'$, and $Y$-axis along $\mib{k}+\mib{k}'$. The $c$- and $a$-axis of the crystal was set parallel to the $Z$- and $Y$-axis, respectively. 
The incident X-ray beam from the synchrotron source is $\pi'$-polarized in the horizontal plane. 
The 006 reflection of a pyrolytic graphite crystal, with $2\theta_{\text{A}}=93.47^{\circ}$ at 7.614 keV, was used as an analyzer. 
By rotating the analyzer angle $\phi_{\text{A}}$ we investigate the polarization state of the diffracted beam. At $\phi_{\text{A}}=0^{\circ}$ and at $90^{\circ}$, only the $\sigma'$ and $\pi'$ polarization component, respectively, of the diffracted beam is measured.  }
\label{fig:ScattConfig}
   \end{minipage}
\hfill
   \begin{minipage}[b]{0.47\textwidth}
\begin{center}
\includegraphics[width=8.5cm]{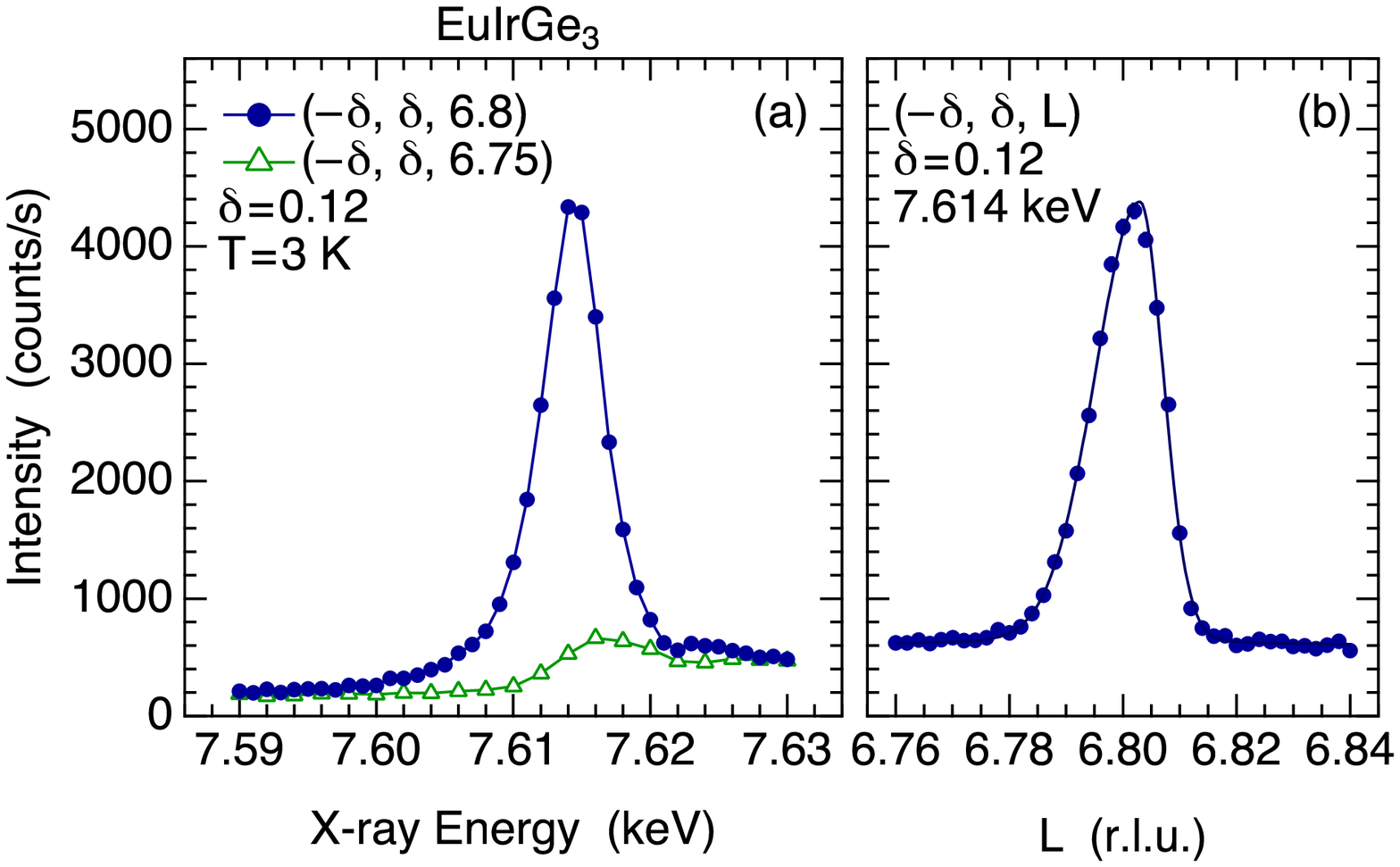}
\end{center}
\caption{(Color online) (a) X-ray energy dependence of the $(-\delta, \delta, 6.8)$ Bragg peak at 3 K. The triangles are the background data taken at $L=6.75$. 
(b) The reciprocal space scan along $(-\delta, \delta, L)$ at 3 K at the $E1$ resonance energy of 7.614 keV, indicating that the peak is exactly at the commensurate position $L=6.8$.  
}
\label{fig:Escan1}
   \end{minipage}
\end{fullfigure}

\begin{fullfigure}
\begin{center}
\includegraphics[width=14cm]{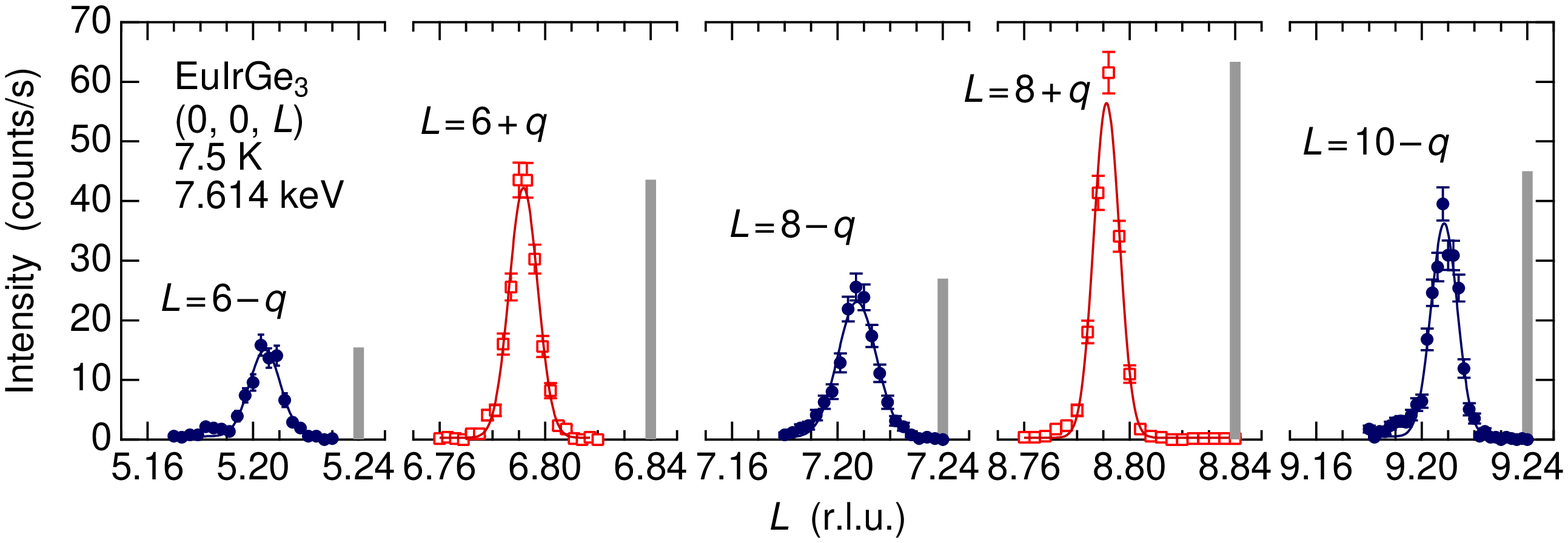}
\caption{(Color online) $L$-scans at 7.5 K in phase I with polarization analysis at $\pi$-$\sigma'$ for $L=6 \pm q$, $8\pm q$, and $10-q$. 
The bars on the right side of the peak show $(A \sin \theta \pm B)^2/\sin 2\theta$ with $A=10$ and $B=0.8$, where $A \sin \theta$ represents the $\pi$-$\sigma'$ structure factor for the $E1$-$E1$ resonance due to magnetic dipole and $B$ represents that for the $E1$-$E2$ resonance, although the phase and the $\theta$ dependence in $B$ are neglected. $\sin 2\theta$ is the Lorentz factor. 
}
\label{fig:LscansWide}
\end{center}
\end{fullfigure}

\begin{fullfigure}
\begin{center}
\includegraphics[width=14cm]{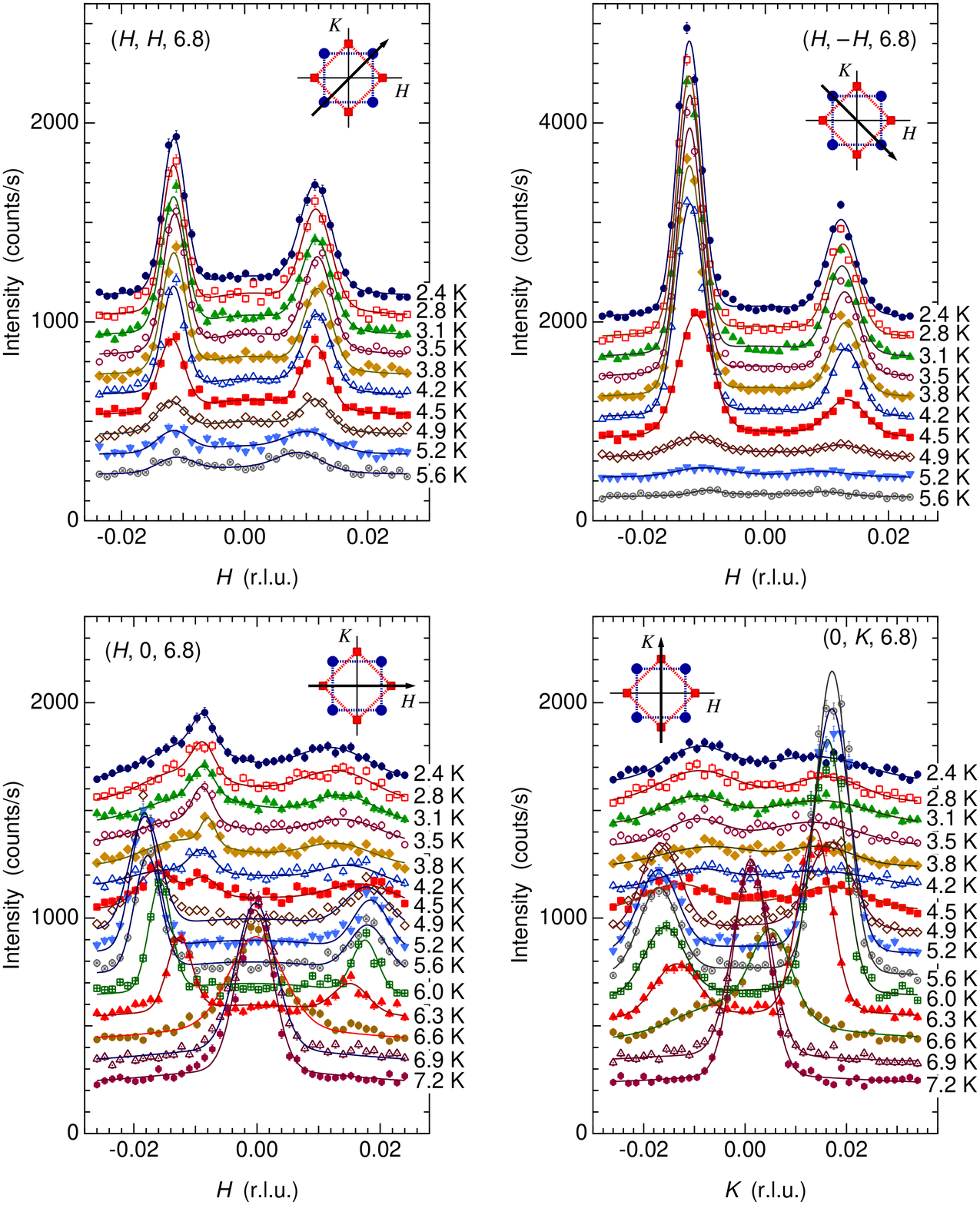}
\caption{(Color online) All of the reciprocal space scans for the $T$-dependence plot of Fig.~\ref{fig:TdepQInt} in the main text. 
The inset figures show the schematic of the peak positions and the ridge structures. The scan line is shown by the arrow. 
}
\label{fig:HKscansAll}
\end{center}
\end{fullfigure}

\clearpage
\begin{figure}
\begin{center}
\includegraphics[width=8.5cm]{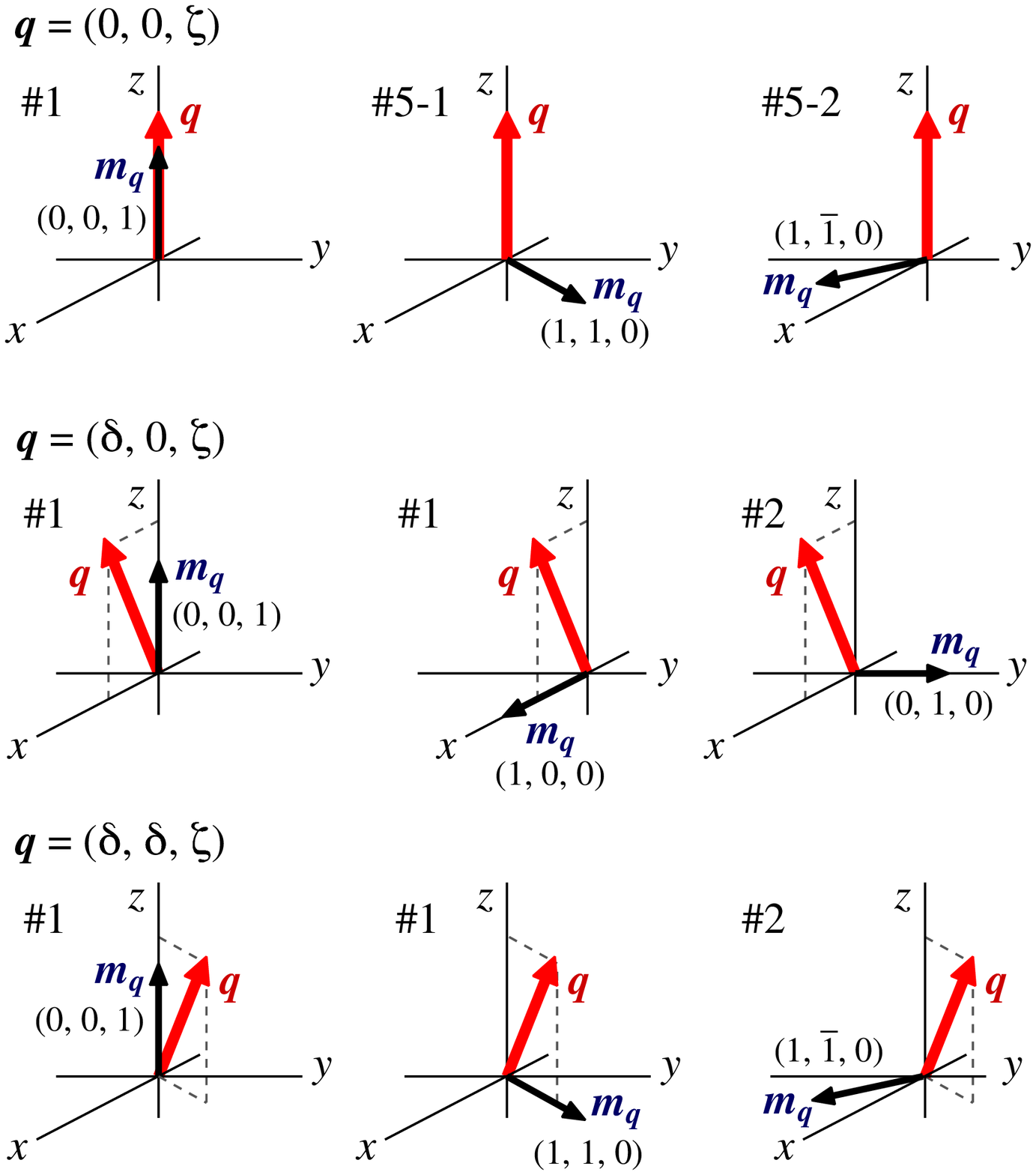}
\end{center}
\caption{(Color online) Irreducible representations of the Fourier component $\mib{m}_{\mib{q}}$ of the magnetic structure of Eu described by a propagation vector $\mib{q}$ in the space group $I4mm$, which were obtained by using the program TSPACE developed by A. Yanase. \#1 and \#2 are the one dimensional representations and \#5 is a two dimensional representation. 
For $\mib{q}=(\delta, 0, \zeta)$ and $(\delta, \delta, \zeta)$, the two components belong to the same representation \#1, allowing the two components to occur simultaneously to form a cycloidal structure. 
}
\label{fig:IrrepI4mm}
\end{figure}

\begin{figure}[h]
\vspace{20mm}
\begin{center}
\includegraphics[width=8.5cm]{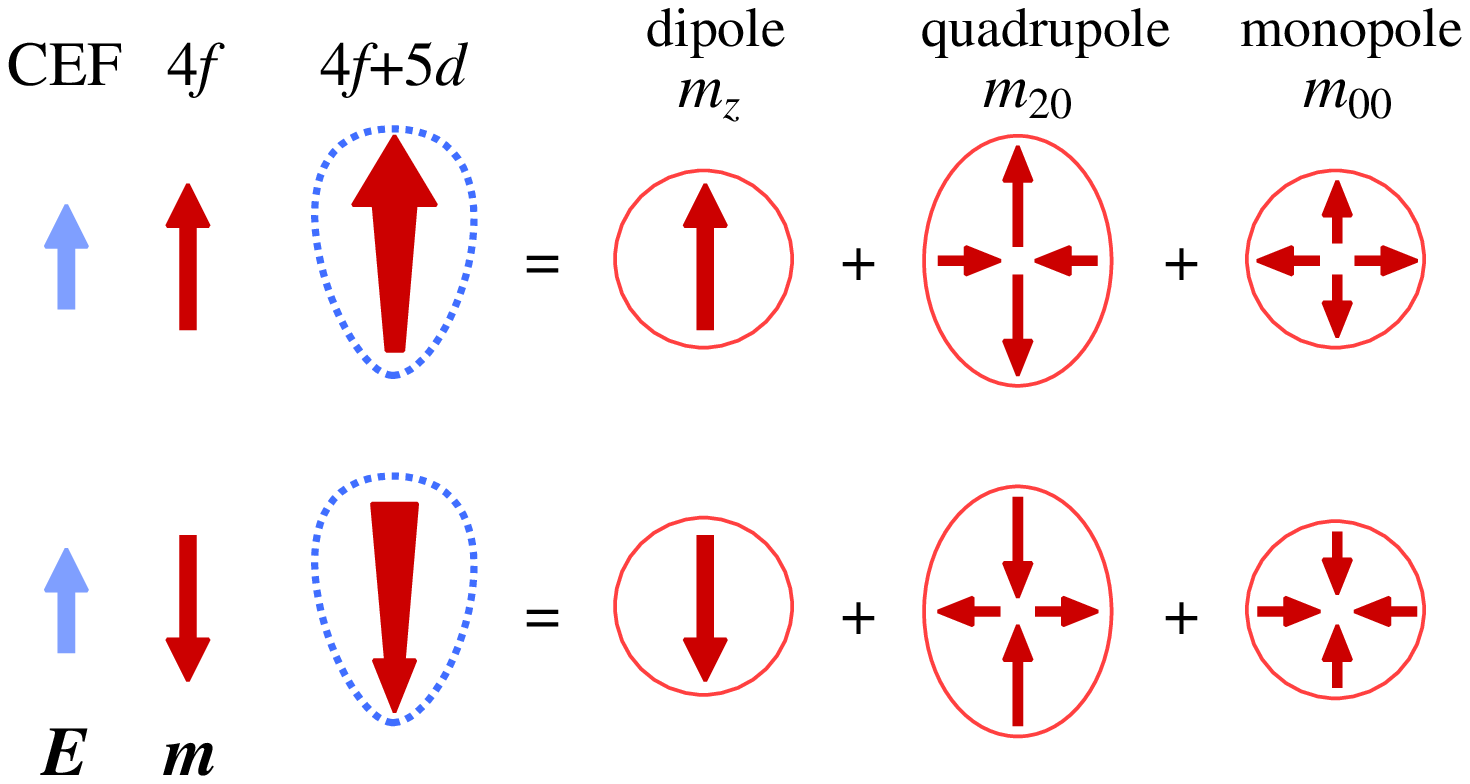}
\end{center}
\caption{(Color online) Schematic of the appearance of odd parity magnetic moments simultaneously with the $4f$ magnetic dipole order in the polar crystal field environment without an inversion symmetry. 
Dotted line represents the polarized charge density of the $4f$+$5d$ hybrid orbital, in which the magnetic moment distribution also lacks the inversion symmetry. 
This state is represented by the superposition of magnetic dipole, magnetic quadrupole, and magnetic monopole moments. 
The even parity dipole moments give rise to an $E1$-$E1$ resonance and the odd parity magnetic quadrupole moments give rise to an $E1$-$E2$ cross-term resonance. 
The different intensity between $(0, 0, G+q)$ and $(0, 0, G-q)$ shown in Fig. S3 arises from the interference between the two processes. 
}
\label{fig:OddParity}
\end{figure}

\newpage
\begin{figure}
\begin{center}
\includegraphics[width=8cm]{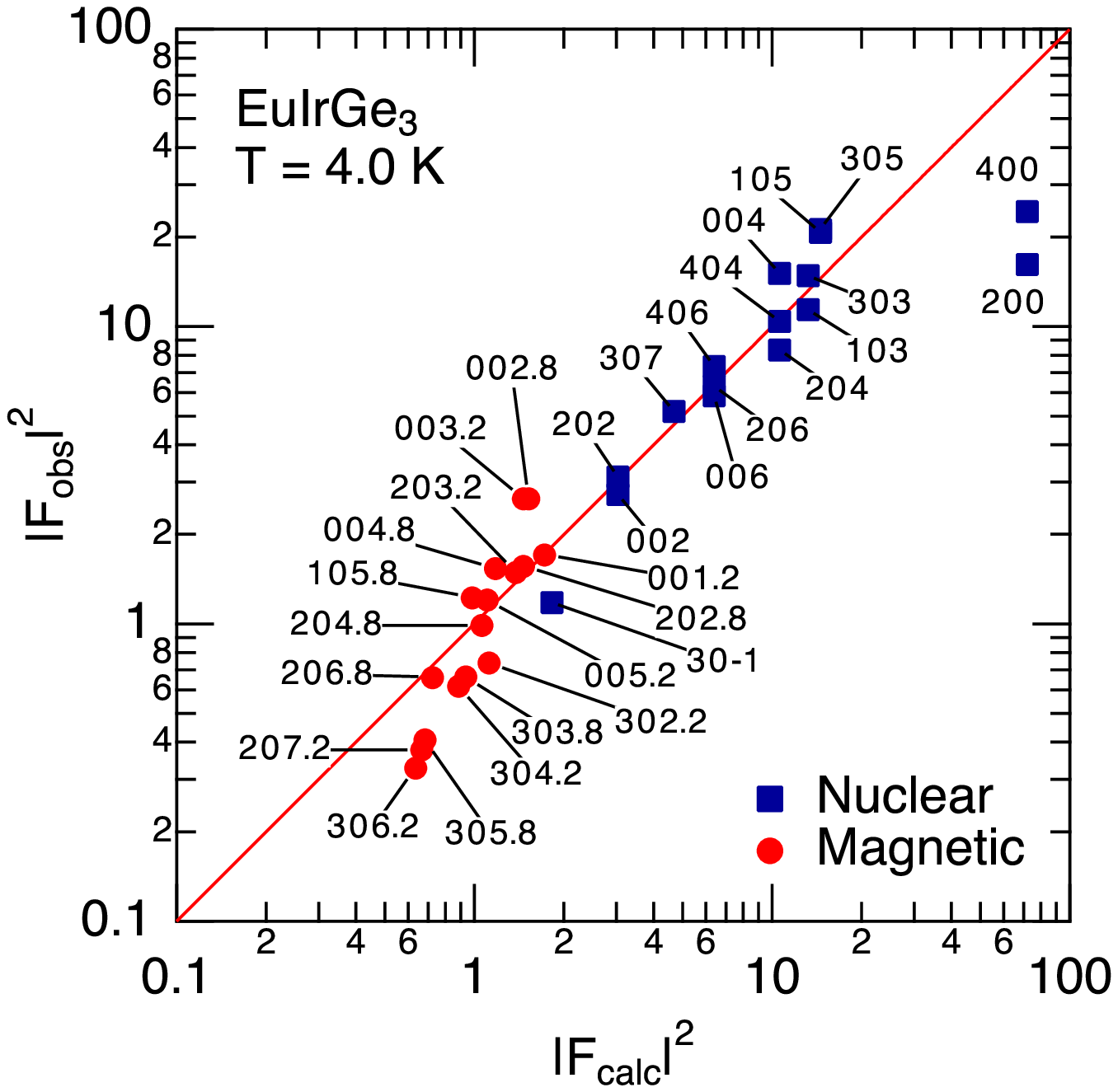}
\caption{(Color online) Comparison of the observed and calculated structure factors for the neutron diffraction measurement at 4.0 K in phase III. 
The observed intensities have been corrected for the Lorentz factor and the absorption; the observed intensity was simply multiplied by the absorption coefficient, which was assumed to be proportional to the inverse of the neutron velocity ($\mu \propto 1/v$). 
The indices of the magnetic reflections represent the center of the four split peaks. We took the sum of all the intensities around the index, including the four peaks and the ridge. The data are compared with the calculation assuming the cycloidal magnetic structure shown in Fig.~\ref{fig:Magst}(c) with the amplitude of 7 $\mu_{\text{B}}$ for the Eu magnetic moment. 
Although the calculation roughly agrees with the observation and we can say that the Eu moment is well developed, the precise determination of the ordered moment is difficult because of the rough estimate of the absorption.  
}
\label{fig:F2obscalc}
\end{center}
\end{figure}


\begin{thebibliography}{99}
\bibitem{Dzyaloshinsky58} I. Dzyaloshinsky, J. Phys. Chem. Solids \textbf{4}, 241 (1958). 
\bibitem{Moriya60} T. Moriya, Phys. Rev. \textbf{120}, 91 (1960). 
\bibitem{Muhlbauer09} S. Muhlbauer, B. Binz, F. Jonietz, C. Pfleiderer, A. Rosch, A. Neubauer, R. Georgii, and P. B\"{o}ni, Science \textbf{323}, 915 (2009). 
\bibitem{Neubauer09} A. Neubauer, C. Pfleiderer, B. Binz, A. Rosch, R. Ritz, P. G. Niklowitz, and P. B\"{o}ni, Phys. Rev. Lett. \textbf{102}, 186602 (2009). 
\bibitem{Kakihana18} M. Kakihana, D. Aoki, A. Nakamura, F. Honda, M. Nakashima, Y. Amako, S. Nakamura, T. Sakakibara, M. Hedo, T. Nakama, and Y. \={O}nuki, J. Phys. Soc. Jpn. \textbf{87}, 023701 (2018). 
\bibitem{Kakihana19} M. Kakihana, D. Aoki, A. Nakamura, F. Honda, M. Nakashima, Y. Amako, T. Takeuchi, H. Harima, M. Hedo, T. Nakama, and Y. \={O}nuki, J. Phys. Soc. Jpn. \textbf{88}, 094705 (2019). 
\bibitem{Kaneko19} K. Kaneko, M. D. Frontzek, M. Matsuda, A. Nakao, K. Munakata, T. Ohhara, M. Kakihana, Y. Haga, M. Hedo, T. Nakama, and Y. \={O}nuki, J. Phys. Soc. Jpn. \textbf{88}, 013702 (2019). 
\bibitem{Tabata19} C. Tabata, T. Matsumura, H. Nakao, S. Michimura, M. Kakihana, T. Inami, K. Kaneko, M. Hedo, T. Nakama, and Y. \={O}nuki, J. Phys. Soc. Jpn. \textbf{88}, 093704 (2019). 
\bibitem{Homma19} Y. Homma, M. Kakihana, Y. Tokunaga, M. Yogi, M. Nakashima, A. Nakamura, Y. Shimizu, D. X. Li, A. Maurya, Y. J. Sato, F. Honda, D. Aoki, Y. Amako, M. Hedo, T. Nakama, and Y. \={O}nuki, J. Phys. Soc. Jpn. \textbf{88}, 094702 (2019). 
\bibitem{Sakakibara19} T. Sakakibara, S. Nakamura, S. Kittaka, M. Kakihana, M. Hedo, T. Nakama, and Y. Onuki, J. Phys. Soc. Jpn. \textbf{88}, 093701 (2019). 
\bibitem{Takeuchi19} T. Takeuchi, M. Kakihana, M. Hedo, T. Nakama, and Y. Onuki, J. Phys. Soc. Jpn. \textbf{88}, 053703 (2019). 
\bibitem{Kurumaji19} T. Kurumaji, T. Nakajima, M. Hirschberger, A. Kikkawa, Y. Yamasaki, H. Sagayama, H. Nakao, Y. Taguchi, T. Arima, and Y. Tokura, Science \textbf{365}, 914 (2019). 
\bibitem{Hirschberger20} M. Hirschberger, T. Nakajima, M. Kriener, T. Kurumaji, L. Spitz, S. Gao, A. Kikkawa, Y. Yamasaki, H. Sagayama, H. Nakao, S. Ohira-Kawamura, Y. Taguchi, T. Arima, and Y. Tokura, Phys. Rev. B \textbf{101}, 220401 (2020). 
\bibitem{Hirschberger19} M. Hirschberger, T. Nakajima, S. Gao, L. Peng, A. Kikkawa, T. Kurumaji, M. Kriener, Y. Yamasaki, H. Sagayama, H. Nakao, K. Ohishi, K. Kakurai, Y. Taguchi, X. Yu, T. Arima, and Y. Tokura, Nature communications \textbf{10}, 5831 (2019). 
\bibitem{Khanh22} N. D. Khanh, T. Nakajima, S. Hayami, S. Gao, Y. Yamasaki, H. Sagayama, H. Nakao, R. Takagi, Y. Motome, Y. Tokura, T. H. Arima, and S. Seki, Adv. Sci. e2105452 (2022). 
\bibitem{Kumar12} N. Kumar, P. K. Das, R. Kulkarni, A. Thamizhavel, S. K. Dhar, and P. Bonville, J. Phys.: Condens. Matter \textbf{24}, 036005 (2012). 
\bibitem{Kaczorowski12} D. Kaczorowski, B. Belan, and R. Gladyshevskii, Solid State Commun. \textbf{152}, 839 (2012). 
\bibitem{Kakihana17} M. Kakihana, H. Akamine, K. Tomori, K. Nishimura, A. Teruya, A. Nakamura, F. Honda, D. Aoki, M. Nakashima, Y. Amako, K. Matsubayashi, Y. Uwatoko, T. Takeuchi, T. Kida, M. Hagiwara, Y. Haga, E. Yamamoto, H. Harima, M. Hedo, T. Nakama, and Y. \={O}nuki, J. Alloys and Compounds \textbf{694}, 439 (2017). 
\bibitem{Rai21} B. K. Rai, G. Pokharel, H. S. Arachchige, S.-H. Do, Q. Zhang, M. Matsuda, M. Frontzek, G. Sala, V. O. Garlea, A. D. Christianson, and A. F. May, Phys. Rev. B \textbf{103}, 014426 (2021). 
\bibitem{Goetsch13} R. J. Goetsch, V. K. Anand, and D. C. Johnston, Phys. Rev. B \textbf{87}, 064406 (2013). 
\bibitem{Maurya14} A. Maurya, P. Bonville, A. Thamizhavel, and S. K. Dhar, J Phys Condens Matter \textbf{26}, 216001 (2014). 
\bibitem{Ryan16} D. H. Ryan, J. M. Cadogan, R. Rejali, and C. D. Boyer, J. Phys.: Condens. Matter \textbf{28}, 266001 (2016). 
\bibitem{Fabreges16} X.  Fabr\`{e}ges, A. Gukasov, P. Bonville, A. Maurya, A. Thamizhavel, and S. K. Dhar, Phys. Rev. B \textbf{93}, 214414 (2016). 
\bibitem{Iha20} W. Iha, S. Matsuda, M. Kakihana, D. Aoki, A. Nakamura, M. Nakashima, Y. Amako, T. Takeuchi, M. Kimata, Y. Otani, M. Hedo, T. Nakama, and Y. \={O}nuki, JPS Conf. Proc. \textbf{30}, 011092 (2020). 
\bibitem{Bednarchuk15} O. Bednarchuk and D. Kaczorowski, Acta Phys. Pol. A \textbf{127}, 418 (2015). 
\bibitem{Maurya16}  A. Maurya, P. Bonville, R. Kulkami, A. Thamizhavel, and S. K. Dhar, J. Magn. Magn. Mater. \textbf{401}, 823 (2016). 
\bibitem{Utsumi18}  Y. Utsumi, D. Kasinathan, P. Swatek, O. Bednarchuk, D. Kaczorowski, J. M. Ablett, and J.-P. Rueff, Phys. Rev. B \textbf{97}, 115155 (2018). 
\bibitem{Momma11} K. Momma and F. Izumi, J. Appl. Crystallogr. \textbf{44}, 1272 (2011). 
\bibitem{Ohhara16} T. Ohhara, R. Kiyanagi, K. Oikawa, K. Kaneko, T. Kawasaki, I. Tamura, A. Nakao, T. Hanashima, K. Munakata, T. Moyoshi, T. Kuroda, H. Kimura, T. Sakakura, C.-H. Lee, M. Takahashi, K. Ohshima, T. Kiyotani, Y. Noda, and M. Arai, J. Appl. Crystallogr. \textbf{49}, 120 (2016). 
\bibitem{SM} (Supplemental Material) The following pieces of information are provided online. 
(S1) Scattering configuration of the resonant X-ray diffraction experiment.  
(S2) X-ray energy dependence and the $L$-scan data at 3 K in phase III.
(S3) $L$-scans at 7.5 K in phase I around $L=6\pm q$, $8\pm q$, and $10-q$. 
(S4) All the scan data for the $T$-dependence plot of Fig.~\ref{fig:TdepQInt}. 
(S5) Irreducible representation of the Fourier component of the magnetic structure. 
(S6) Schematic of the appearance of odd parity magnetic moments with the magnetic dipole order. 
(S7) Comparison of the observed and calculated structure factors for the neutron diffraction measurement at 4.0 K in phase III. 
\bibitem{Hannon88} J. P. Hannon, G. T. Trammell, M. Blume, and D. Gibbs: Phys. Rev. Lett. \textbf{61} (1988) 1245; \textbf{62} (1988) 2644.
\bibitem{Blume94} M. Blume: in \textit{Resonant Anomalous X-ray Scattering, Theory and Applications}, ed. G. Materlik, C. J. Sparks, and K. Fischer (Elsevier Science, Amsterdam, 1994) p. 495.
\bibitem{Lovesey05} S. W. Lovesey, E. Balcar, K. S. Knight, and J. Fern\'{a}ndez-Rodr\'{i}guez, Phys. Rep. \textbf{411}, 233 (2005). 
\bibitem{Kaneko21} K. Kaneko, T. Kawasaki, A. Nakamura, K. Munakata, A. Nakao, T. Hanashima, R. Kiyanagi, T. Ohhara, M. Hedo, T. Nakama, and Y. \={O}nuki, J. Phys. Soc. Jpn. \textbf{90}, 064704 (2021). 
\bibitem{Shimomura19} S. Shimomura, H. Murao, S. Tsutsui, H. Nakao, A. Nakamura, M. Hedo, T. Nakama, and Y. \={O}nuki, J. Phys. Soc. Jpn. \textbf{88}, 014602 (2019). 
\bibitem{Shang21} T. Shang, Y. Xu, D. J. Gawryluk, J. Z. Ma, T. Shiroka, M. Shi, and E. Pomjakushina, Phys. Rev. B \textbf{103}, L020405 (2021). 
\bibitem{Kezsmarki15} I. K\'{e}zsm\'{a}rki, S. Bord\'{a}cs, P. Milde, E. Neuber, L. M. Eng, J. S. White, H. M. R{\o}nnow, C. D. Dewhurst, M. Mochizuki, K. Yanai, H. Nakamura, D. Ehlers, V. Tsurkan, and A. Loidl, Nat Mater \textbf{14}, 1116 (2015). 
\bibitem{Hayami18a} S. Hayami and H. Kusunose, J. Phys. Soc. Jpn. \textbf{87}, 033709 (2018). 
\bibitem{Hayami18b} S. Hayami, M. Yatsushiro, Y. Yanagi, and H. Kusunose, Phys. Rev. B \textbf{98}, 165110 (2018). 
\bibitem{Yatsushiro21} M. Yatsushiro, H. Kusunose, and S. Hayami, Phys. Rev. B \textbf{104}, 054412 (2021). 
\end{thebibliography}
\end{document}